\documentclass[twocolumn,aps,prl]{revtex4}
\usepackage{amsmath}
\usepackage{graphicx}

\begin{document}

\title{Heat Transfer between Weakly Coupled Systems: Graphene on {\rm a}-${\rm SiO_2}$}
\author{ B.N.J. Persson$^{1,2}$ and H. Ueba$^{1}$}

\affiliation{$^1$ Division of Nanotechnology and New Functional Material Science,
Graduate School of Science and Engineering,
University of Toyama, Toyama, Japan}
\affiliation{$^2$IFF, FZ-J\"ulich, 52425 J\"ulich, Germany, EU}

\begin{abstract}
We study the heat transfer between weakly coupled systems with flat interface.
We present a simple analytical result which can be used to estimate the heat transfer coefficient.
As an application we consider the heat transfer between graphene and amorphous ${\rm SiO_2}$. 
The calculated value of the heat transfer coefficient is in good agreement 
with the value deduced from experimental data.
\end{abstract}

\maketitle

\pagestyle{empty}


Almost all surfaces in Nature and Technology have roughness on many different length scales\cite{PSSR}.
When two macroscopic solids are brought into contact, even if the applied force is very small,
e.g., just the weight of the upper solid block, the pressure in the asperity contact regions 
can be very high, usually close to the yield stress of the (plastically) softer solid.
As a result good thermal contact may occur within each microscopic contact region, but owing
to the small area of real contact the (macroscopic) heat transfer coefficient may still be small.
In fact, recent studies have shown that in the case of surfaces with roughness on many different
length scales, the heat transfer is {\it independent}
of the area of real contact\cite{PLV}. We emphasize that this remarkable and counter-intuitive result is only valid when
roughness occur over several decades in length scale.

For nanoscale systems the situation may be very different. 
Often the surfaces are very smooth with typically nanometer
(or less) roughness on micrometer-sized surface areas, 
and because of adhesion the solids often make contact over a large fraction of the
nominal contact area. 
The heat transfer between solids in perfect contact is usually calculated using
the so called diffusive mismatch model\cite{Pohl}, where it is assumed that all phonons scatter diffusively and
elastically at the interface between two materials. In this model there is no direct reference to
the nature of the solid-solid interaction accross the interface, and the model cannot describe
the heat flow between weakly interacting solids.

Here we will discuss the heat transfer across perfectly flat interfaces, when the
interaction between the solids is very weak, e.g., of the Van der Waals type, as for graphene or
carbon nanotubes on many substrates. We present a simple 
analytical result which can be used to estimate the heat transfer coefficient.
As an application we consider the heat transfer between graphene and amorphous ${\rm SiO_2}$. 
The calculated value of the heat transfer coefficient is in good agreement 
with the value deduced from experimental data.

\begin{figure}[tbp]
\includegraphics[width=0.4\textwidth,angle=0]{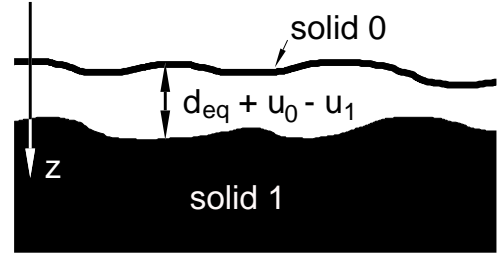}
\caption{
A membrane (solid {\bf 0}) in contact with a semi-infinite solid {\bf 1}. The interfacial
surface separation is the sum of the equilibrium separation $d_{\rm eq}$ and the difference 
in the surface displacements $u_0-u_1$, due to thermal movements, 
where both  $u_0$ and  $u_1$  are positive when the displacement point
along the  $z$-axis towards the interior of solid {\bf 1}.} 
\label{picshem}
\end{figure}

Consider the interface between two solids, and assume that local thermal equilibrium occurs everywhere except at the 
interface. The energy flow (per unit area) through the interface is given by\cite{PLV}
$$J= \alpha (T_0-T_1),$$
where $T_0$ and $T_1$ are the local temperatures at the interface in solid {\bf 0} and {\bf 1}, respectively.
We now present a calculation of the heat transfer coefficient 
$\alpha$ under the assumption of very weak coupling between the two solids.

The stress or pressure acting on the surface of solid ${\bf 1}$ from solid ${\bf 0}$ can be written as
$$\sigma ({\bf x},t) = K[u_0({\bf x},t)-u_1({\bf x},t)],$$
where $u_0$ and $u_1$ are the (perpendicular) surface displacement of solid ${\bf 0}$ and  ${\bf 1}$ 
(see Fig. \ref{picshem}), respectively, and where  $K$
is a spring constant per unit area characterizing the interaction between the two solids.
If we define
$$u ({\bf q},\omega)  = {1\over (2\pi )^{3}} \int d^2x dt \ u({\bf x},t) e^{-i({\bf q}\cdot {\bf x} -\omega t)}, $$
we get
$$\sigma ({\bf q},\omega) = K[u_0({\bf q},\omega)-u_1({\bf q},\omega)].\eqno(1)$$
If solid ${\bf 1}$ is semi-infinite we have\cite{PJCP}
$$u_1 ({\bf q},\omega) = M({\bf q},\omega) \sigma ({\bf q},\omega), \eqno(2)$$
where  $M({\bf q},\omega)$ is determined by the elastic properties of solid ${\bf 1}$.
Combining (1) and (2) gives
$$u_1 ({\bf q},\omega) = {K M({\bf q},\omega) \over 1+K M({\bf q},\omega)} u_0 ({\bf q},\omega). \eqno(3)$$
The energy transferred to solid ${\bf 1}$ from solid ${\bf 0}$ during the time period $t_0$ can be written as
$$\Delta E = \int d^2 x d t \  \dot u_1({\bf x},t) \sigma ({\bf x},t),$$
where  $\dot u = \partial u/ \partial t$. One can also write 
$$\Delta E =(2\pi )^3 \int d^2 q d \omega \  (-i\omega) u_1({\bf q},\omega) \sigma (-{\bf q},-\omega).\eqno(4)$$
Using (3) and (4) we obtain
$$\Delta E = (2\pi )^3 \int d^2 q d \omega \  {\omega  K^2 {\rm Im} M({\bf q},\omega) \over |1+K M({\bf q},\omega)|^2} 
\langle |u_0({\bf q},\omega)|^2\rangle, \eqno(5)  $$
where we have performed an ensemble (or thermal) average denoted by  $ \langle .. \rangle $.
Next, note that
$$\langle |u_0({\bf q},\omega)|^2\rangle =
{A_0 t_0 \over (2\pi )^3} C_{uu}({\bf q},\omega), \eqno(6)$$
where $A_0$ is the surface area, and 
$$C_{uu}({\bf q},\omega) =  {1\over (2\pi )^{3}} \int d^2x dt \ \langle u_0({\bf x},t) u_0(0,0)\rangle e^{i({\bf q}\cdot{\bf x}-\omega t)}, $$
is the displacement correlation function.
We can also write
$$C_{uu}({\bf q},\omega) =   
\int d^2 q' d \omega' \  
\langle u_0(-{\bf q},-\omega) u_0({\bf q}',\omega')\rangle . \eqno(7)  $$
Substituting (6) in (5) gives the heat current $J_{0\rightarrow 1} = \Delta E/ A_0 t_0$ from solid  ${\bf 0}$ to  solid ${\bf 1}$:
$$J_{0\rightarrow 1} 
= 2 \int d^2 q \int_0^\infty d \omega \ {\omega  K^2 {\rm Im} M({\bf q},\omega)\over |1+K M({\bf q},\omega)|^2}  
C_{uu}({\bf q},\omega;T_0).  \eqno(8)$$
where $C_{uu}$ depends on the temperature  $T_0$ of solid  ${\bf 0}$.
A similar equation with $T_0$ replaced by $T_1$ gives the energy transfer from solid ${\bf 1}$ to solid ${\bf 0}$, and the
net energy flow $J=J_{0\rightarrow 1}-J_{1\rightarrow 0}$.

With graphene in mind, we 
now assume that solid ${\bf 0}$ is a membrane or two-dimensional (2D) system. We assume that $u_0({\bf x},t)$ satisfies
$$ \rho_0 {\partial^2 u_0 \over \partial t^2} = -\kappa \nabla^2 \nabla^2 u_0 -\rho \eta {\partial u_0 \over \partial t} +f, \eqno(9)$$
where $\kappa \approx 1 \ {\rm eV}$ \cite{Fasolino}  is the bending elasticity, $\eta$ 
a phenomenological friction coefficient, and $f({\bf x},t)$
a stochastic fluctuating force related to the temperature $T_0$ and the friction  $\eta$ via the fluctuation-dissipation theorem:
$$\langle  f({\bf q},\omega)  f({\bf q}',\omega') \rangle = 2 (2\pi)^{-3} \rho \eta k_{\rm B}T_0 
\delta (\omega+\omega') \delta ({\bf q}+{\bf q}').\eqno(10)$$
where $\rho_0 = n_0 m_0$ is the mass density per unit area of the 2D-system ($m_0$ is the atom mass and $n_0$ the number of
atoms per unit area).
On the right hand side of (9) there should in principle be another term $K \nabla^2 u_0$, due to the interaction with the substrate
wall, but because of the assumed small magnitude of $K$, this term can be neglected for the relevant wavevectors $q$. 
The formalism above assumes that the system can be treated classically which is the case only if $\hbar \omega << k_{\rm B} T_0$. If this
condition is not satesfied one can take into account the most important quantum mechanical effects by replacing $k_{\rm B} T_0$
by $\Pi (\omega)= \hbar \omega \left [{\rm exp} (\hbar \omega / k_{\rm B} T_0) - 1\right ]^{-1/2}.$

From (9) we get
$$u_0({\bf q},\omega) = {f({\bf q},\omega) \over -\rho \omega^2 +\kappa q^4 -\rho \eta i \omega}$$
Using (7) and (10) we get
$$C_{uu} ({\bf q},\omega) = 
{ 2(2\pi)^{-3} \rho \eta k_{\rm B}T_0  \over |\rho \omega^2 -\kappa q^4 +\rho \eta i \omega |^2}.$$
In the limit $\eta \rightarrow 0$ we get
$$C_{uu} ({\bf q},\omega)  = {\pi \over (2\pi)^3} {1 \over \rho_0 \omega_1^2}\delta (\omega-\omega_1),\eqno(11) $$
where $\omega_1 = (\kappa /\rho_0)^{1/2} q^2 = c(q) q$, where we have defined the velocity $ c(q)= (\kappa /\rho_0)^{1/2} q$.  

Substituting (11) in (8) and assuming weak coupling between the solids (i.e., $K$ is small), we get
$$J_{0\rightarrow 1}  
 =  {k_{\rm B} T_0\over 2 \pi \rho_0}  \int_0^\infty dq \  {q\over \omega_1}  K^2 {\rm Im} M({\bf q},\omega_1) . $$
The heat transfer coefficient $\alpha = (J_{0\rightarrow 1}-J_{1\rightarrow 0})/(T_0-T_1)$ is
given by
$$ \alpha =  {k_{\rm B} \over 2 \pi \rho_0}  \int_0^\infty dq \  {q\over \omega_1}  K^2 {\rm Im} M({\bf q},\omega_1) .\eqno(12)$$
Using the expression for $M({\bf q}, \omega)$ derived in \cite{PJCP,Ryberg} and $\omega_1 = c(q) q$ gives
$$ \alpha =  {k_{\rm B} K^2 \xi \over  \rho_0 \rho_1 c_{\rm T}^3}, \eqno(13)$$
where
$$ \xi =  {1 \over 2 \pi}  \int_0^{q_{\rm c}} dq \  {1\over q} {c(q)\over c_{\rm T}} $$ 
$$ \times {\rm Re} \left( \left [{c^2(q)\over c_{\rm L}^2}-1\right ]^{1/2} \over 
\left [{c^2(q)\over c_{\rm T}^2}-2\right ]^2 +4\left [{c^2(q)\over c_{\rm T}^2}-1 \right ]^{1/2} 
\left [{c^2(q)\over c_{\rm L}^2}-1\right ]^{1/2} \right ), $$
where $c_{\rm L}$, $c_{\rm T}$ and $\rho_1$ are the longitudinal and transverse 
sound velocities, and the mass density, respectively, of solid ${\bf 1}$.
The cut off wavevector $q_{\rm c} \approx \pi /a_1$ ($a_1$ is the lattice constant, or the average distance between two 
nearby atoms) of solid {\bf 1}.

There are two contributions to the integral $\xi$. One is derived from $c(q) > c_{\rm L}$,
but for graphene on a-${\rm SiO_2}$ studied below, this gives only $\sim 10\%$ of the contribution to the integral.
For $c(q) < c_{\rm L}$ the term after the Re operator is purely imaginary (and will therefore not contribute
to the integral), {\it except} for the case where the denominator vanish. It is found that
this pole-contribution gives the main contribution ($\sim 90\%$) to the integral, and corresponds to the excitation of
a Rayleigh surface (acoustic) phonon of solid {\bf 1}. This process involves 
energy exchange between a bending vibrational mode 
of the graphene and a Rayleigh surface phonon mode of solid {\bf 1}. The denominator vanish when $c(q)=c_{\rm R}$ where
$$\left [{c^2_{\rm R}\over c_{\rm T}^2}-2\right ]^2 - 4\left [1-{c^2_{\rm R}\over c_{\rm T}^2}\right ]^{1/2} 
\left [1-{c^2_{\rm R}\over c_{\rm L}^2}\right ]^{1/2} =0 . $$
Note that the 
Rayleigh velocity $c_{\rm R} < c_{\rm T}$ but close to $c_{\rm T}$. 
For example, when $c_{\rm L}/ c_{\rm T} = 2$,  $c_{\rm R} \approx 0.93 c_{\rm T}$,
and the pole contribution to the integral in $\xi$ is $0.083$.

In the model above the heat transfer between the solids involves a single bending mode of the
membrane or 2D-system. In reality there will always be some roughness at the interface which will somewhat
blurre the wavevector conservation rule. We therefore expect some narrow band of bending modes to be involved in
the energy transfer, rather than a single mode. Nevertheless, the model study above 
assumes implicitly that, due to lattice non-linearity (and defects), there exist phonon scattering 
processes which rapidly transfer energy to the bending mode
involved in the heat exchange with the substrate. 
This requires very weak coupling to the substrate, so that the
energy transfer to the substrate is so slow that the bending mode can be re-populated by phonon scattering 
processes in the 2D-system, e.g., from the in-plane phonon modes,
in such a way that its population is always close to what would be the case if complete thermal equilibrium 
occurs in the 2D-system. This may require high temperature in order for multi-phonon
scattering processes to occur by high enough rates. 

Graphene, the recently isolated 2D-carbon material with unique properties due to its linear electronic
dispersion, is being actively explored for electronic applications\cite{Geim}. 
Important properties are the high mobilities reported especially in suspended graphene, the fact that graphene is the ultimately thin
material, the stability of the carbon-carbon bond in graphene, the ability to induce a bandgap by electron confinement in graphene
nanoribbons, and its planar nature, which allows established pattering and etching techniques to be applied.
Recently it has been found that the heat generation in graphene field-effect transistors can result 
in high temperature and
device failure\cite{Freitag}. Thus, it is important to understand the the mechanisms which influence the heat flow. 

\begin{figure}[tbp]
\includegraphics[width=0.4\textwidth,angle=0]{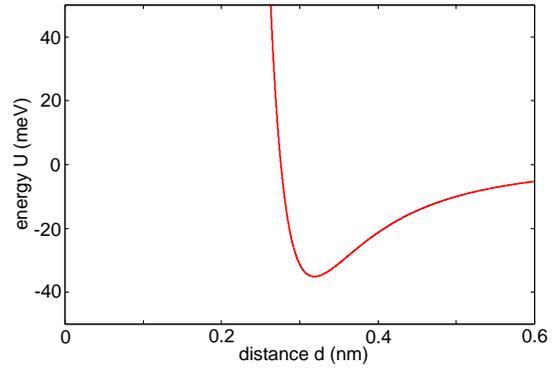}
\caption{
The calculated ${\rm graphene}-{\rm a}$-${\rm SiO_2}$ interaction energy $U(d)$ per graphene carbon atom, as a function
of the separation $d$ (in nm) between the center of a graphene carbon atom and the center of the first layer
of substrate atoms. See text for details.}
\label{d.nm.U.meV.per.C.atom.graphene.SiO2}
\end{figure}

The ${\rm graphene}-{\rm a}$-${\rm SiO_2}$ interaction is probably of the van Der Waals type. In Ref. \cite{Pop} the interaction between
the graphene C-atoms and the substrate Si and O atoms was assumed to be described by the Lennard-Jones (LJ) pair-potentials with different
parameters. Here we use a simplified picture where the substrate atoms form a simple cubic lattice with the lattice constant
determined by $a_1=(\bar m/\rho_1)^{1/3} \approx 0.25 \ {\rm nm}$, 
where $\bar m = (m_{\rm Si}+2 m_{\rm O})/3 \approx 3.32\times 10^{-26} \ {\rm kg}$ is the average substrate atomic mass,
and $\rho_1\approx 2200 \ {\rm kg/m^3} $ the mass density of a-${\rm SiO}_2$. We also use the effective LJ energy
parameter, $\epsilon = (\epsilon_{\rm Si}+2\epsilon_{\rm O})/3\approx 5.3 \ {\rm meV}$, and the bond-length parameter 
$\sigma = (\sigma_{\rm Si}+2\sigma_{\rm O})/3 \approx 0.31 \ {\rm nm}$. With these parameters we can calculate 
the ${\rm graphene}-$a-${\rm SiO_2}$ interaction energy, $U(d)$, per graphene carbon atom, as a function
of the separation $d$ (in nm) between the center of a graphene carbon atom and the center of the first layer
of substrate atoms. We find (see Fig. \ref{d.nm.U.meV.per.C.atom.graphene.SiO2}) the
${\rm graphene}-$a-${\rm SiO_2}$ binding energy $E_{\rm b} = -U(d_{\rm eq}) \approx \ 35 \ {\rm meV}$ 
per carbon atom, and the force constant $k=U''(d_{\rm eq})$
(where $d_{\rm eq}\approx 0.32 \ {\rm nm}$ is the equilibrium separation)
$k= K a_0^2 = 2.4 \ {\rm N/m}$ per carbon atom. This gives the perpendicular ${\rm graphene}-{\rm a}$-${\rm SiO_2}$ (uniform) vibration
frequency $\omega_\perp \approx (k/m_0)^{1/2} \approx \ 55 \ {\rm cm}^{-1}$, 
which is similar to what is observed for the perpendicular 
vibrations of linear alkane molecules on many surfaces (e.g., about $ 55-60 \ {\rm cm}^{-1}$ for alkanes on
metals and on hydrogen terminated diamond C(111)\cite{Woell}). 
Using $K = k/ a_0^2 = 1.82\times 10^{20} \ {\rm N/m^3}$, and the transverse and longitudinal sound velocities of
solid {\bf 1} ($c_{\rm T} = 3743 \ {\rm m/s}$ and $c_{\rm L} = 5953 \ {\rm m/s}$), from (13) we obtain 
$\alpha \approx 3\times 10^8 \ {\rm W/Km^2}$. 

The heat transfer coefficient between graphene and a {\it perfectly flat} a-${\rm SiO_2}$ substrate has not been measured
directly, but measurements of the heat transfer between carbon nanotubes and sapphire 
by Maune et al\cite{Maune} indicate that it may be of order $\alpha \approx 8\times 10^8  \ {\rm W/m^2K}$.
This value was deduced indirectly by measuring the breakdown voltage of carbon nanotubes, which could be related to the
temperature increase in the nanotubes. Molecular dynamics calculations\cite{Pop} for nanotubes on a-${\rm SiO_2}$  gives 
$\alpha \approx 3 \times 10^8  \ {\rm W/m^2K} $ (here it has been assumed that the contact width between the nanotube and
the substrate is $1/5$ of the diameter of the nanotube). 
Finally, using a so called 3 $\omega $ method, Chen et al\cite{Chen} have measured the heat 
transfer coefficient $\alpha \approx 2 \times 10^8 \ {\rm W/m^2 K}$.

As pointed out in Sec. 3, the model developed 
above for the heat transfer involves a single, or narrow band, of bending modes of the
membrane or 2D-system. In order for this model to be valid, the coupling to the substrate must be so weak that the
energy transfer to the substrate from the bending mode occur so slowly that the mode can be re-populated by phonon scattering 
processes, in such a way that its population is always close to what is expected if full thermal equilibrium 
would occur within the 2D-system. This may require high temperature in order for multi-phonon
scattering processes to occur by high enough rates. We suggest that this may be one reason for the 
decrease in the heat transfer coefficient observed for the  ${\rm graphene}-{\rm a}$-${\rm SiO_2}$ system 
below room temperature\cite{Chen}. 

In Ref. \cite{Freitag} the temperature profile in the graphene under current was 
studied experimentally. The heat transfer coefficient between graphene 
and the {\rm a}-${\rm SiO_2}$ substrate was determined by modeling the heat flow using the standard
heat flow equation with the heat transfer coefficient as the only unknown quantity. The authors found that using
a constant (temperature independent) heat transfer coefficient  $\alpha \approx 2.5\times 10^7 \ {\rm W/m^2K}$
the calculated temperature profiles in graphene are in good agreement with experiment. This $\alpha $ is at least
10 times smaller than expected for perfectly flat surfaces (see above).
In an earlier paper\cite{PUeba} we have studied the heat transfer between graphene and a-${\rm SiO_2}$.
In that study we assumed that because of surface roughness 
the graphene only makes partial contact with the  ${\rm SiO_2}$ substrate,
which will reduce the heat transfer coefficient as compared to the perfect contact case.
We showed how observed heat transfer can be explained as resulting 
from the spreading resistance, rather than the heat resistance from the area of real contact.

It has recently been suggested\cite{Rotkin} that the heat transfer between graphene and a-${\rm SiO_2}$ may involve
photon tunneling\cite{rev1}. That is, coupling via the electromagnetic field between electron-hole pair
excitations in graphene and optical phonons in a-${\rm SiO_2}$. However, our calculations indicate that
for graphene adsorbed on  a-${\rm SiO_2}$ the field coupling gives a negligible contribution to the heat transfer\cite{PUeba}.

To summarize, we have studied the heat transfer between weakly coupled systems with flat interface.
We have presented simple analytical results which can be used to estimate the heat transfer coefficient.
Detailed results was presented for the heat transfer between a membrane and a semi-infinite solid. For this case
the energy transfer is dominated by energy exchange between a bending vibrational mode 
of the graphene, and a Rayleigh surface phonon mode of the substrate. 
This  model assumes implicitly that, due to lattice non-linearity (and defects), there exist phonon scattering 
processes which rapidly transfer energy to the bending mode
involved in the heat exchange with the substrate. 
This may require high temperature in order for multi-phonon
scattering processes to occur at high enough rate.
As an application we have considerd the heat transfer between graphene and amorphous ${\rm SiO_2}$. 
The calculated value of the heat transfer coefficient was found to be in good agreement 
with the value deduced from the experimental data.

\vskip 0.2cm
We thank P. Avouris for drawing our attention to Ref. \cite{Freitag}. 
B.N.J.P. was supported by Invitation Fellowship Programs for Research in Japan from
Japan Society of Promotion of Science (JSPS).
H.U. was supported  by the Grant-in-Aid for Scientific
Research B (No. 21310086) from JSPS.


\begin{thebibliography}{99}

\bibitem{PSSR}
B.N.J. Persson, Surf. Sci. Rep. {\bf 61}, 201 (2006).

\bibitem{PLV}
B.N.J. Persson, B. Lorenz and A.I. Volokitin, The European Physics Journal E{\bf 31}, 3 (2010).

\bibitem{Pohl}
E.T. Swartz and R.O. Pohl, Rev. Mod. Phys. {\bf 61}, 605 (1989).

\bibitem{PJCP}
B.N.J. Persson, Journal of Chemical Physics {\bf 115}, 3840 (2001).

\bibitem{Fasolino}
A. Fasolino, J.H. Los and M.I. Katsnelson,
NATURE {\bf 6}, 858 (2007).

\bibitem{Ryberg}
B.N.J. Persson and R. Ryberg, Phys. Rev. B{\bf 32}, 3586 (1985).

\bibitem{Geim}
A.K. Geim and K.S. Novoselov, Nat. Mater. {\bf 6}, 183 (2007).

\bibitem{Freitag}
M. Freitag, M. Steiner, Y. Martin, V. Perebeinos, Z. Chen, J.C. Tsang and P. Avouris,
Nano Letters {\bf 9}, 1883 (2009).

\bibitem{Pop}
Z-Y Ong and E. Pop, Phys. Rev. B{\bf 81}, 155408 (2010).

\bibitem{Woell}
M. Fuhrmann and Ch. W\"oll, New Journal of Physics {\bf 1}, 1 (1998).

\bibitem{Maune}
H. Maune, H-Y Chiu and M. Bockrath,
Applied Physics Letter {\bf 89}, 013109 (2006).

\bibitem{Chen}
Z. Chen, W. Jang, W. Bao, C.N. Lau and C. Dames,
Applid Physics Letters {\bf 95}, 161910 (2009).

\bibitem{PUeba}
B.N.J. Persson and H. Ueba, subm. to Nano Letters.

\bibitem{Rotkin}
S. V. Rotkin, V. Perebeinos, A.G. Petrov and P. Avouris,
Nano Letters {\bf 9}, 1850 (2009); M. Freitag, M. Steiner, Y. Martin, V. Perebeinos, Z. Chen, J.C. Tsang and P. Avouris, 
Nano Letters {\bf 9}, 1883 (2009).

\bibitem{rev1}
A.I. Volokitin and B.N.J. Persson, Reviews of Modern Physics {\bf 79}, 1291 (2007).

\end{thebibliography}
\end{document}